# EPHEMERIS RECONSTRUCTION FOR COMET 67P/CHURYUMOV-GERASIMENKO DURING ROSETTA'S PROXIMITY PHASE FROM RADIOMETRIC DATA ANALYSIS

Riccardo Lasagni Manghi,* Marco Zannoni,† Paolo Tortora‡, Frank Budnik§, Bernard Godard**, Nicholas Attree††

This study provides a continuous ephemeris reconstruction for comet 67P/Churyumov-Gerasimenko by reanalyzing Rosetta radiometric measurements and Earth-based astrometry. Given the comet-to-spacecraft relative trajectory provided by the navigation team, these measurements were used to estimate the comet state and some critical physical parameters, most notably the non-gravitational accelerations induced by the outgassing of surface volatiles, for which different models were tested and compared. The reference reconstructed ephemeris, which uses a stochastic acceleration model, has position uncertainties below 10 km, 30 km, and 80 km in the orbital radial, tangential, and normal directions for the whole duration of the Rosetta proximity phase (from July 2014 and October 2016). Furthermore, the solution can fit ground-based astrometry between March 2010 and July 2018, covering a complete heliocentric orbit of 67P. The orbital radial and normal components dominate the estimated comet's non-gravitational accelerations, reaching peak values of $(1.28 \pm 0.17)\cdot 10^{-8}$ m/s$^2$ and $(0.52 \pm 0.20)\cdot 10^{-8}$ m/s$^2$, respectively 15 days and 24 days after perihelion. Furthermore, the acceleration magnitude is shown to have a steep dependence on the comet's heliocentric distance ($NGA \sim r^{-6}$) and shows asymmetries in the pre- and post- perihelion activities. The estimated acceleration components, which are agnostic due to the limited physical assumptions, could be used as a constraint for future investigations involving high-fidelity thermophysical models of the comet's surface.

## INTRODUCTION

Accurate estimation of cometary trajectories has proven challenging due to the presence of non-gravitational accelerations (NGA) induced by the sublimation of sub-surface volatiles. Given the extreme complexity of the outgassing process, these accelerations appear stochastic, causing short-term and secular variations to the comets' orbital parameters that limit the validity of ephemeris prediction over successive perihelion passes.

* Junior Assistant Professor, Dipartimento di Ingegneria Industriale, Alma Mater Studiorum - Università di Bologna, Bologna, Italy
† Senior Assistant Professor, Dipartimento di Ingegneria Industriale, Alma Mater Studiorum - Università di Bologna, Bologna, Italy.
‡‡ Full Professor, Dipartimento di Ingegneria Industriale, Alma Mater Studiorum - Università di Bologna, Bologna, Italy.
§ European Space Agency, ESA-ESOC, Darmstadt, Germany.
** European Space Agency, ESA-ESOC, Darmstadt, Germany.
†† Post-doctoral researcher, Instituto de Astrofisica de Andalucia (IAA-CSIC), Granada, Spain

After an early characterization by Whipple[1], a wide array of models has been proposed to describe these perturbing accelerations. Most of these models are very complex and depend on poorly known parameters linked to the surface's chemical composition and thermophysical properties[2,3] or the morphology of the comet's nucleus[4].

For practical orbit determination applications and short-term prediction of cometary motion, it is often helpful to rely on simple empirical models such as the one proposed by Marsden[5], hereby referred to as the standard model (SM), where the NGA is expressed in the orbital radial, transverse, and normal (RTN) reference frame as the product of constant scale factors $A_i$, representing the normalized acceleration components at 1 AU, and a function of the comet's heliocentric distance $g(r) \propto r^{-k}$, related to the water ice sublimation rate (see Equation 1).

$$\vec{a}_{SM} = g(r)(A_R \hat{e}_R + A_T \hat{e}_T + A_N \hat{e}_N) \tag{1}$$

A generalization of this model is provided by the extended standard model (ESM)[6], which introduces a time offset between the perihelion and the peak of outgassing activity. In this formulation, the function $g(r)$ in Equation (1) is replaced by $g(r')$, where $r' = r(t') = r(t - \Delta T)$, and $\Delta T$ is the time delay between the perihelion and the peak of outgassing activity.

In the more recent rotating jet model (RJM)[7], the accelerations are expressed as the sum of a discrete number of jets localized on the comet's surface, whose contributions are averaged over a single rotational period, as shown in Equation (2). In this expression, n is the number of jets, $A_{J_i}$ is the intrinsic jet strength (i.e., the equivalent acceleration at 1AU assuming the Sun is at the local Zenith), $\Delta\theta$ is the diurnal lag angle, $\hat{e}_P$ is the direction of the comet's pole, $\hat{e}_S$ is the most sunward direction in the comet's equatorial plane, and $\hat{e}_Q$ completes the right-handed frame. The parameters $J_P$ and $J_S$, whose values depend on the season and the jet's colatitude $\eta_i$ (i.e., the angle that it forms with the North pole), account for the average insolation of the jet over a rotational period. Their complete formulation is given by Chesley & Yeomans[7].

$$\vec{a}_{RJM} = g(r') \sum_{i=1}^{n} A_{J_i} \left( J_S \cos\Delta\theta \, \hat{e}_R + J_S \sin\Delta\theta \, \hat{e}_Q + J_P \hat{e}_P \right)_i \tag{2}$$

By accounting for seasonal variations of the NGA, the rotating jet model is particularly suited for reconstructing cometary ephemerides spanning multiple orbital periods or using high-precision observational data[8]. Although this model is highly flexible, it may suffer from a loss of predictive power when the data don't provide enough resolution to constrain the additional solved-for parameters.

Another model explored in this study is a linear stochastic model, for which the acceleration components in the orbital RTN frame are defined as piecewise continuous linear accelerations for each stochastic interval. The expression for each component is given in Equation (3), where $\vec{\alpha}_k$ is the initial acceleration vector at the $k^{th}$ stochastic interval, $\vec{\beta}_k$ is the acceleration derivative (jerk), and $t_k$ is the initial interval epoch. Continuity of the acceleration values at the boundary of successive intervals was enforced through the constraints in Equation (4), where the length of each interval is defined as a function of the comet's heliocentric distance as $T_k = min\,(T_0 \cdot (r'/r_0)^5, \Delta T_{max})$. In this expression, the minimum length of the stochastic interval at perihelion $T_0$ and the maximum length of the interval $\Delta T_{max}$ represent the free parameters of the problem. This approach was preferred over using constant stochastic intervals as it allowed for a better representation of the short-time variability of NGA around the perihelion.

$$\vec{\alpha}_k(\mathrm{t}) = \vec{\alpha}_k + (t - t_k) \cdot \vec{\beta}_k \tag{3}$$

$$\vec{\alpha}_{k+1} - \left(\vec{\alpha}_k + T_k \cdot \vec{\beta}_k\right) \ll 1 \tag{4}$$

To mitigate the effect of measurement biases due to relative orbit discontinuities and produce more realistic estimations of the NGA, a series of loose constraints were added between the successive stochastic accelerations. The expression for these constraints is given in Equation (5), where $\sigma_{\beta_k}$ is the uncertainty of the jerk component at the $k^{th}$ stochastic interval, and $C$ represents a variable constraint factor, which is an additional free parameter. Finally, the estimated acceleration and jerk values at the first and last stochastic intervals are constrained to be null to ensure zero acceleration values at high heliocentric distances ($R$ > 3.5 $AU$).

$$\beta_{k+1} - \beta_k < \frac{1}{C} \cdot \left(\sigma_{\beta_k}^2 + \sigma_{\beta_{k+1}}^2\right)^{\frac{1}{2}}\Bigg|_{R,T,N} \tag{5}$$

Regardless of the employed dynamical model, accurate ephemeris reconstruction and estimation of the NGAs acting on cometary nuclei as a function of time around perihelion represent critical steps towards understanding the physical mechanisms involved in the outgassing process. The Rosetta mission represents a unique opportunity, thanks to its prolonged proximity operations at 67P/Churyumov-Gerasimenko (67P/CG) during its active period.

Rosetta is an ESA cornerstone mission whose main objective was to rendezvous with the short-period comet 67P/CG and to monitor the evolution of the comet's activity as it followed its heliocentric trajectory. Launched in 2004, Rosetta endured a 10-year cruise phase, performing four gravity assists with the Earth and Mars and two asteroid flybys at 2876 Steins and 21 Lutetia. Following its arrival at 67P/CG on August 6th, 2014, Rosetta successfully navigated the comet's proximity for more than two years, using a combination of radiometric measurements and optical images collected by the onboard navigation cameras[9,10,11]. During this period, Rosetta delivered the first lander to ever touch down on a cometary surface (Philae) and collected a wide array of in-situ measurements before ending its mission with a controlled collision on 30th September 2016[12,13]. By escorting 67P/CG as it passed through its perihelion, Rosetta provided a unique opportunity to investigate the comet's non-gravitational motion induced by surface outgas- sing, thanks to the availability of continuous and accurate radiometric measurements.

Rosetta performed several orbit determination analyses as part of the navigation process, which required periodic updates of the comet's ephemeris and rotational states. However, due to limited knowledge of the physical properties of the nucleus, outgassing-induced NGAs were not included in the dynamical model. As a result, close to the perihelion, where a peak of outgassing was expected, the comet's ephemeris needed more frequent updates from a rate of roughly once every 1-2 weeks up to once every three days during critical mission phases. The resulting heliocentric trajectory for 67P/CG, obtained by merging the various operational solutions generated by the Flight Dynamics and Flight Control Teams at ESOC (hereby referred to as ESOC FD), contains several discontinuities at the boundaries of these orbit determination intervals. These discontinuities have magnitudes as high as 10-100 km in the J2000 inertial reference frame, which correspond to geocentric range discontinuities in the order of a few hundred meters around perihelion, as shown in Figure 1.

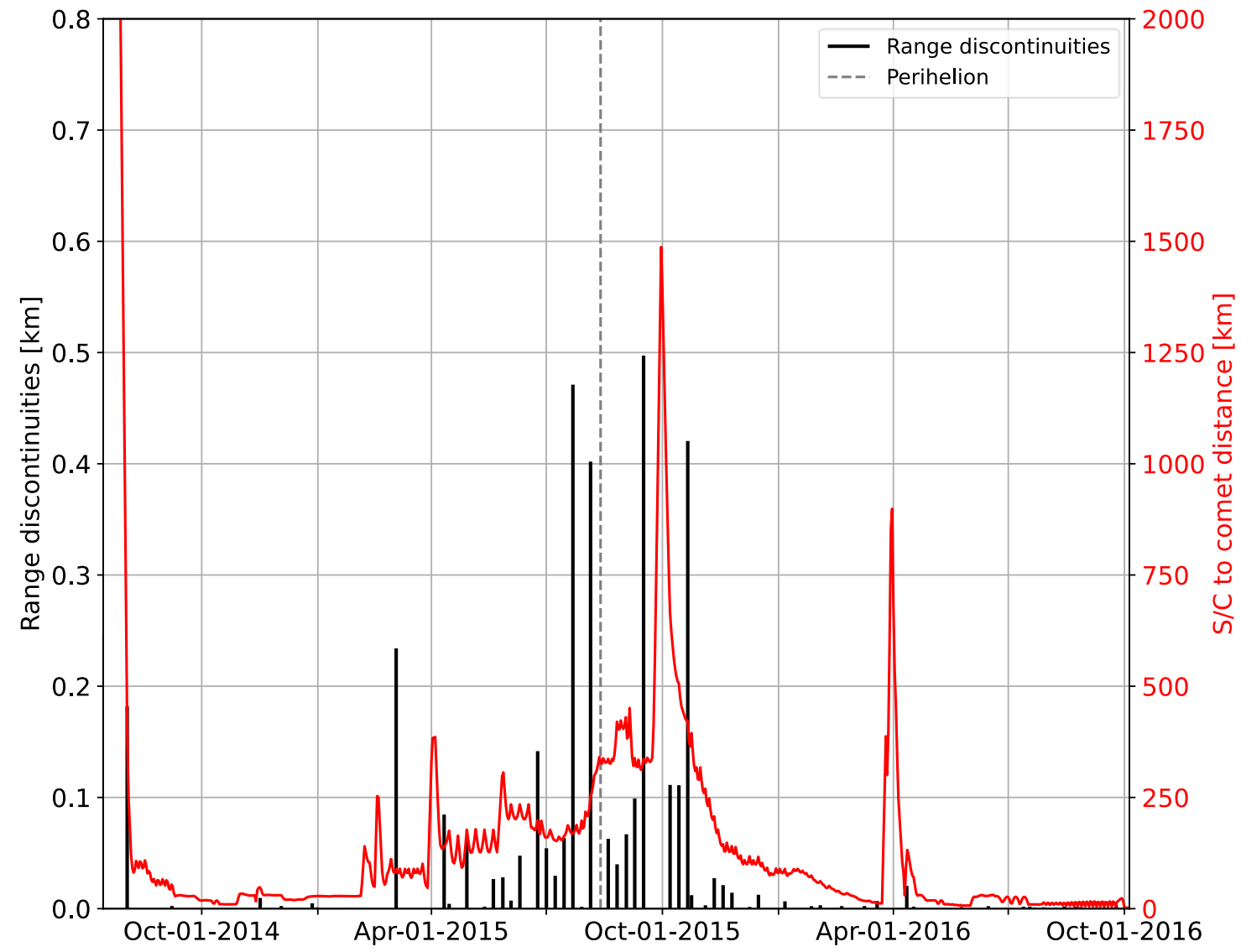

**Figure 1 Geocentric range discontinuities for 67P/CG (ESOC FD solution) as a function of the spacecraft to comet distance.**

A new ephemeris solution for 67P/CG was realized in a recent study by JPL[8], which combines ranging data collected during the Rosetta mission with ground-based astrometric measurements to reconstruct the comet's trajectory from 2012 to 2018, covering an entire orbital period. This study employed a restricted number of range observables collected when the spacecraft was close to the comet, which are highly accurate thanks to the tight constraint on the relative spacecraft trajectory provided by the Doppler and optical navigation measurements. Furthermore, only high-accuracy astrometric measurements collected before and after the most active outgassing periods were used, while those around the time of perihelion passage were discarded to avoid potential biases induced by the comet's tail. For these reasons, the study produced a robust orbit reconstruction for the covered period with high predictive power for the comet ephemeris. However, using a global outgassing model and restricting the dataset around perihelion prevents the resolution of short time-scale variations of the NGAs, which represents a crucial step towards linking the observed non-gravitational motion to the physical properties of the comet's surface.

This study provides a continuous and accurate ephemeris reconstruction for comet 67P/CG between July 2014 and September 2016, along with its associated uncertainty. Strictly coupled with the ephemeris reconstruction is the estimation of the NGA induced by surface outgassing, representing an additional product of the analysis. To fulfill these objectives, the current investigation uses radiometric measurements collected during Rosetta's proximity phase, including measurements around perihelion, and ground-based astronomy data retrieved from the Minor Planet Center (MPC)[*].

## METHODS

The high-level orbit determination approach for this analysis is similar to the one used for the ephemerides reconstruction of planets and small bodies using deep-space probes and planetary orbiters[14,15]. The original range measurements are usually reduced to "normal points" by adjusting for relevant parameters, including the spacecraft's relative orbit, the target's mass, and its gravity field. The resulting residuals represent derived corrections to the original ephemerides used for the reduction. As such, when added to the instantaneous geocentric range, they provide a pseudo-range observable between the Earth's center and the center of the target[16]. These pseudo-range observables are then processed along with radar range and astrometric measurements to estimate the ephemerides of the target bodies.

[*] https://www.minorplanetcenter.net/

For the current analysis, no measurement reduction is performed. Radiometric measurements, referenced to the spacecraft's center, are combined with Rosetta's relative trajectory around the comet to estimate the cometary ephemerides directly. Specifically, the employed relative trajectory is a combination of the orbit determination reconstructions realized by the navigation team, which were merged in a single Spice kernel after the end of the mission. Consistent with the approach described above, the spacecraft-referenced radiometric measurements were de-weighted to account for Rosetta's relative orbit uncertainty. Since the formal uncertainty of the relative orbit reconstruction is not available at the time of the current study, an empirical weighting scheme was developed, which links the orbit uncertainty to the instantaneous cometocentric distance (see Equation 6 below). The approach used to model the non-gravitational dynamics relies on simple empirical formulations and aims to provide an agnostic estimation of the NGA, which could be used as a reference for higher-fidelity outgassing models as part of further studies[17].

The OD analysis was performed using a batch least-square filter implemented in NASA-JPL's orbit determination program MONTE (Mission Analysis, Operations, and Navigation Toolkit Environment), currently used for the operations of all NASA's space missions managed by JPL[18], as well as for radio science data analysis and processing[19,20]. MONTE's mathematical formulation and measurement models are detailed in Moyer[21,22].

**Data selection and processing**

During Rosetta's proximity phase, radiometric measurements at X-band (8.5 GHz) were acquired by ESA's ESTRACK complexes of Cebreros, Malargüe, and New Norcia and by NASA's DSN complexes of Canberra, Goldstone, and Madrid. In this analysis, range and ΔDOR measurements collected between July 2014 and September 2016, which are publicly available through the Planetary Science Archive[*], were used. It should be mentioned that all ΔDOR observables were recorded within narrow time intervals in July 2014 and February 2016. To provide an additional constraint to the comet's plane of the sky position, radiometric observables were complemented by the Earth-based astrometric measurements publicly available on the MPC at the time of this study (July 2024).

The radiometric observables were corrected for Earth media propagation delays. Specifically, ionospheric calibrations were derived from GNSS dual-frequency measurements[23]. Tropospheric calibrations for DSN complexes were similarly derived from GNSS measurements. In contrast, tropospheric calibrations for ESTRACK stations were computed using local meteorological data for the dry component of the delay and seasonal models for the wet component[24].

As part of the data processing routines, range observables were reduced by selecting a single representative measurement for each tracking pass at a given station, corresponding to the observable recorded at the highest elevation. The standard deviation associated with the reduced measurements is the root square sum of the systematic and random errors of the original range measurements. Having typical values around 5 m, after calibration of station and media delays, range biases are roughly one order of magnitude higher than typical noise values of [50, 70] $cm$ for a single arc and are, therefore, representative of the whole pass. Furthermore, from the set of reduced measurements, we retained a single range point for each orbit determination arc of the ESOD FD solution to avoid any cross-correlation between the processed datasets.

[*] https://psa.esa.int/psa/#/pages/home

ESOC FD provided ΔDOR observables in the form of ASCII text files, each containing a set of three DOR measurements with a duration of a few minutes, which were taken in a Spacecraft-Quasar-Spacecraft sequence.

Ground-based astrometric observations of 67P were retrieved from the MPC database. Within this dataset, we distinguish between standard measurements and a subset of high-accuracy astrometric measurements from Cerro Paranal, Mt. Lemmon Survey, and Pan-STARRS 1, having sub-milli-arcsecond accuracies. These latter were retrieved by JPL as part of an effort to reconstruct the ephemeris of 67P and were added to the database for public dissemination. The JPL work also highlights that standard MPC data collected around the August 2014 perihelion has significantly higher noise levels and shows clear signatures in the declination residuals, likely due to the asymmetric morphology of the comet tail. For this reason, we restricted the analysis to data collected at higher heliocentric distances ($R > 3$ AU). Furthermore, MPC measurements that reported the magnitude as total show larger residuals with respect to the ones where the magnitude is not reported or related to the nucleus only. Therefore, only data belonging to the former categories were used within the filter. Finally, all astrometric data were corrected to account for possible biases caused by using pre-GAIA stellar catalogs in the image processing[25,26].

A summary of all processed measurements with their corresponding receiving stations and availability intervals is given in Table 1 .

**Table 1 Summary of processed observables. The number of observables for the 2-way range refers to the reduced measurements (one per OD arc of Rosetta). Similarly, MPC astrometry refers to the number of measurements used in the filter after data reduction.**

| Type | Complex / Baseline / Source | Observables | Start | End |
|---|---|---|---|---|
| 2-Way Range | ESTRACK | 87 | 02-Aug-2014 | 29-Sep-2016 |
| | Deep Space Network | 53 | 20-Oct-2014 | 30-Sep-2016 |
| ΔDOR | Cebreros-Malargue | 7 | 24-Jul-2014 | 26-Feb-2016 |
| | Cebreros-New Norcia | 7 | 24-Jul-2014 | 24-Feb-2016 |
| Astrometry | Minor Planet Center | 999 | 08-Mar-2010 | 14-Nov-2014 |
| | JPL (Farnocchia et al. 2021) | 169 | 13-Mar-2014 | 14-Jun-2018 |

### Data weights

For the orbit determination approach described above, the uncertainty of the radiometric observables is a function of the intrinsic measurement noise, the relative orbit uncertainty, and the geometry of the relative orbit with respect to the Earth-comet direction. The standard approach for planetary ephemerides is to select a fixed weight for the pseudo-range normal points, corresponding to the characteristic orbit uncertainty over the period of interest[27,14,15]. This approach works well for planetary orbiters, where the relative distance between the target and the probe is kept more or less constant for prolonged periods. In the case of Rosetta, the accuracy of the relative orbit reconstruction, which primarily relied on optical and Doppler measurements, changed significantly over time due to the variable dynamical environment around 67P/CG (i.e., due to outgassing-induced drag on the spacecraft structures and complex gravitational potential) and the wide range of spacecraft to comet distances for the various mission phases (see Figure 1).

Since the full covariance of the spacecraft ephemeris solution used for navigation was not available at the time of writing, the empirical formulation in Equation (6) was used, which expresses the relative orbit uncertainty as a function of the spacecraft to comet distance R.

$$\sigma(R)\ [km] = \begin{cases} 1.5 \cdot 10^{-4} \cdot R^{1.02} \\ 0.2 \cdot 10^{-4} \cdot R^{1.42} \end{cases} for \begin{cases} R \leq 100\ km \\ R > 100\ km \end{cases} \quad (6)$$

Specifically, the first expression, used for values of R<100 km, was obtained by fitting a power-law to the formal position uncertainties obtained for the long-arc orbit determination solution by ESOC FD (see Figure 11 of Reference 27). The second expression, used for values of R >100 km, was instead obtained by fitting a power law to the geocentric range discontinuities for 67P/CG, which were used as proxies for Rosetta's position uncertainty close to the perihelion, and imposing continuity with the first expression, as shown in Figure 2. The validity of this assumption was confirmed by performing a filter passthrough with the operational trajectories for both Rosetta and 67P/CG and inspecting the range residuals. If the relative spacecraft position was accurately known, one would expect to observe some features corresponding to the comet's trajectory discontinuities. However, these features are not observed, indicating that the geocentric range discontinuities for 67P/CG are compensated for by equal and opposite discontinuities in the relative trajectory, implying that the relative position uncertainty was, at best, comparable with the magnitude of the observed jumps. A more detailed description of the validation process is provided in APPENDIX.

The characteristic distance of 100 km, used in the definition of the uncertainty, represents a conventional transition point between landmark-based optical navigation, where the spacecraft position is estimated using a combination of known surface features (e.g., boulders or craters), and a centroid-based one, where the comet's center of brightness in the pictures is used to estimate the expected center of mass. This distance represents the approximate limit above which the target cannot be spatially resolved within the optical navigation images.

After calibrating station and media delays, 1-way range measurements have typical uncertainties of ~1 m, so the relative orbit uncertainty will dominate the measurement noise. Therefore, 2-way range weights are derived by doubling the uncertainty in Equation (6). Uncertainties of the ΔDOR measurements were computed as the root square sum of the uncertainties for the spacecraft and quasar VLBI measurements, which represented an output of the cross-correlator and were provided by ESOC FD. Typical values for ΔDOR uncertainties are in the order of 0.5 $ns$, which correspond to roughly [3, 6] $km$ in the Earth-spacecraft normal direction at typical geocentric distances encountered during the Rosetta mission. Measurements collected in February 2016, when the spacecraft was in proximity of the comet, are thus dominated by the original measurement uncertainty, while measurements collected in July 2014 during the Far Approach Trajectory phase have uncertainties comparable to the assumed relative orbit uncertainty. Thus, ΔDOR weights were obtained as the root square sum of the original ΔDOR uncertainty and the relative orbit uncertainties computed with Equation (6) at the corresponding measurement epochs.

Measurement weights for the high-accuracy astrometry derived by JPL, computed as the root mean square of the residuals for each observatory, range between roughly 0.1 and 0.4 arcseconds[8]. For the standard MPC astrometry, we employed the weighting scheme proposed by Veres et al., which is based on a statistical analysis of the astrometric errors of the major asteroid surveys[28]. According to this scheme, individual weights are functions of the observatory, the date of the observation, the star catalog used for the image processing, and the number of measurements per observing session from the same station.

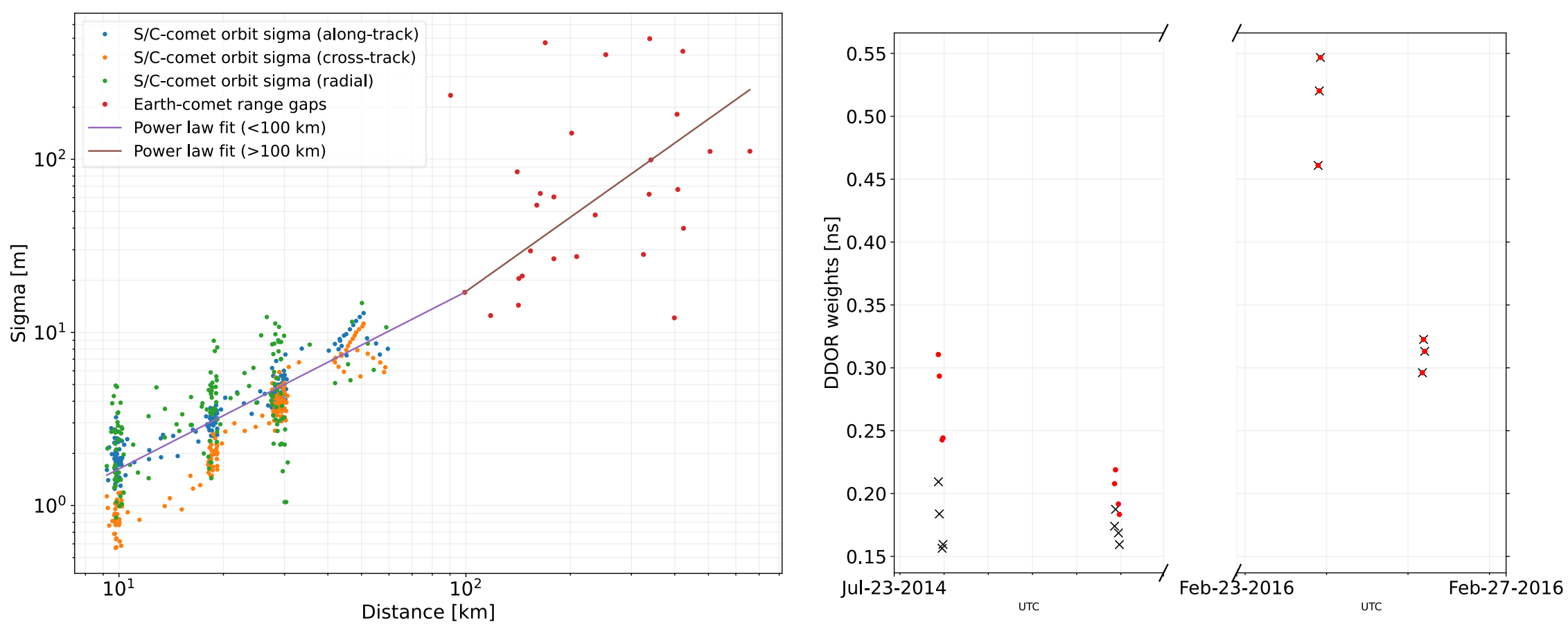

**Figure 2 Left: weighting scheme for the 1-way range measurements; circles represent the formal position covariances from Reference 29 (green, blue, and orange) and the geocentric range discontinuities for 67P/CG (red); continuous lines represent the power law fits for the corresponding range regimes in Equation (6). Right: ΔDOR weights; black crosses: measurement uncertainties from the cross-correlator; red dots: final ΔDOR weight, including the relative orbit uncertainty.**

## Dynamical model

The gravitational accelerations that were considered for this analysis include relativistic point-mass gravity from the Sun, the planets and their satellites, the Moon, and Pluto; Newtonian point-mass gravity from the 343 most massive bodies in the main asteroid belt; gravitational $J_2$ perturbation from the Sun's oblateness. State vectors and gravitational parameters for the Sun, the planets, their satellites, and the small bodies were taken from JPL's DE440.

Non-gravitational accelerations induced by surface outgassing were introduced using several models with increasing levels of complexity, which represent alternative test cases for the analysis.

The first model to be evaluated was the standard model, which represents a baseline for the dynamics of active cometary nuclei. Several test cases were explored by considering different power laws for the sublimation rate $g(r)$ and by including specific acceleration components in the heliocentric RTN frame. The second model to be evaluated was the extended standard model, for which we used the formulation of Yeomans & Chodas[6], which allows us to estimate the time offset $\Delta T$ between the perihelion and the peak of outgassing activity. Since the comet state and its partials are only known for previous times steps, when integrating the comet's trajectory, the partial derivatives of $g(r)$ with respect to the comet's state are determined at each integration step using a two-body trajectory and the osculating orbital elements corresponding to the integration time. This approach, which simplifies the mathematical formulation for the differential corrector, is justified by the observation that estimated time offsets are often small, in the order of a few weeks.

The third model represents a simplified version of the rotating jet model, in which the comet spin pole orientation is known and assumed to be fixed throughout the mission, with values of right ascension $\alpha = 69°$ and declination $\delta = 65°$ derived from previous investigations on 67P/CG[30]. Only two jets were used to model the outgassing, one in the northern and one in the southern hemisphere, each capturing the overall dynamics of the respective portions of the surface.

In addition to the global models described above, a Linear Stochastic Model (LSM) was used, for which the acceleration components in the orbital RTN frame were modeled as piecewise linear functions. This model can rely on minimal assumptions on the physical mechanism causing the NGA, providing an agnostic solution. However, this comes at the cost of increased parameter space, requiring careful selection of the stochastic interval length and the constraints applied to the model's estimated parameters. On one side, short intervals and loose constraints allow for capturing short-scale variations of the NGA components around the perihelion, which are essential to linking the observed non-gravitational motion to the surface physical processes. On the other side, longer stochastic intervals and tighter constraints make the solution more robust against observational biases, such as Rosetta's relative orbit discontinuities, which the estimated parameters might absorb.

### Filter setup

**Table 2** summarizes the estimated parameters that are common to all the analyzed test cases. The following sections will describe additional solved-for parameters related to specific NGA models. *A priori* values for the comet's state were taken from the heliocentric trajectory generated by the navigation team at the reference epoch of 13-Aug-2015 02:04 (ET), corresponding to the perihelion passage time.

A single range bias was estimated locally for each filter arc that included pseudo-noise measurements from the Malargüe station, amounting to a total of 14 parameters. *A priori* values and uncertainties for the pseudo-noise range biases were derived from the passthrough solution generated using the operational trajectories for Rosetta and 67P/CG. Specifically, *a priori* values represented the average offset between pseudo-noise and sequential range measurements within individual arcs, while the uncertainty was taken from the standard deviation of pseudo-noise range residuals.

**Table 2 Estimated parameters and their corresponding *a priori* uncertainties**

| Parameter | N | *A priori* sigma | Comments |
|---|---|---|---|
| Comet position | 3 | $10^4\ km$ | Widely open |
| Comet velocity | 3 | $100\ m/s$ | |
| Pseudo-noise range biases | 14 | $10\ m$ | From the std of pseudo-noise range residuals using the passthrough |

## RESULTS

Several test cases were analyzed as part of the orbit determination process by changing the model used to represent the NGA induced by surface outgassing and changing the *a priori* values and uncertainties for the estimated parameters. To evaluate the ability of a given model to fit the observed data, we used an approach similar to the one employed for reconstructing the trajectory of 1I/2017 U1 'Oumuamua[31], which relies on the evaluation of the reduced chi-square statistics.

### Comparison of NGA models

Table 3 summarizes the statistics of the residuals for the considered test cases. In the first scenario, we considered the standard model. Following previous studies on short-period comets, the normal component of the acceleration was assumed to be significantly smaller than the in-plane components[32,33] and was not considered due to the poor observability along the orbit-normal direction. Therefore, the only solved-for parameters were the comet state at perihelion and the scale factors $A_R$ and $A_T$ from Equation 1, for which an *a priori* uncertainty of $10^{-7} m/s^2$ was used[*].

---

[*] Typical accelerations for short period comets are between $10^{-8}$ m/s$^2$ and $10^{-6}$ m/s$^2$ [31]. Previous analyses for 67P/CG estimated values of $10^{-8}$ m/s$^2$ and $10^{-9}$ m/s$^2$ for the radial and tangential/normal components, respectively[32,33].

Different power laws were used to approximate the water production rate by varying the exponent k between 2 and 6. Although the model is unable to fit the data, we can observe that higher values of the exponent k correspond to improved residuals and reduced $\chi^2$ values. A similar trend is observed for the extended standard model. We can observe from Table 3 that $\chi^2$ values are reduced by almost an order of magnitude when switching from the standard formulation of Marsden ($k \cong 2$) to $k = 5$. This finding is in agreement with previous studies indicating that an $r^{-5}$ power law best approximates the water production rate for 67P/CG[8,34], which was therefore used as a baseline for the following models. Adding a third acceleration component along the orbit-normal direction slightly reduces the $\chi^2$ for both the standard and extended standard models. However, the improved range residuals and goodness of fit come at the expense of the ΔDOR and astrometric measurements, which are significantly degraded and show signatures in the declination.

The third test case to be evaluated is the rotating jet model. For this analysis, we assumed a constant pole orientation and a total of two jets, one located in the northern and one in the southern hemisphere of the comet. To narrow down the parameter space, a broad grid search was initially performed by varying the jet's colatitude $\eta_i$ and the diurnal lag angle $\Delta\theta$ with constant steps while keeping the time offset fixed at $\Delta T = 15.22\ days$, corresponding to the estimated value for the best-fit solution of the ESM. The only estimated parameters were the comet state at perihelion and the intrinsic jets' strength $A_{Ji}$. A narrow grid search was then performed by restricting to colatitudes of $\eta_N \in [40°, 80°]$ and $\eta_S \in [130°, 170°]$ for the northern and southern jets, corresponding to the areas of lower $\chi^2$ values for the broad grid search. In addition to the comet state and the jet strength, both the time offset of peak outgassing activity and the diurnal lag angle were estimated within the filter starting from *a priori* values of $\Delta T = 15.22 \pm 5\ days$ and $\Delta\theta = 10° \pm 5°$. The best-fit solution for the narrow grid search, corresponding to colatitudes of $\eta_N = 45°$ and $\eta_S = 140°$, was then used to perform a final filter run where all the parameters of the rotating jet model were estimated, including the jets' colatitudes, which are poorly observable due to the high degree of non-linearity in the NGA (complete formulation in APPENDIX A of Reference 7).

The results of this solution are summarized in Table 3, where we can observe a significant improvement of the filter performances with respect to the extended standard model, with an order of magnitude reduction in the range residuals and $\chi^2_\nu$ values reduced by two orders of magnitude. However, the rotating jet model can still not produce an acceptable fit, indicating that shorter time-scale variations of the NGA values around perihelion might be required to fit the observed measurements well.

**Table 3 Statistics of residuals for the various NGA models.**

| Model | $k$ | $RMS_{range}$ $[m]$ | $RMS_{\Delta DOR}$ $[ns]$ | $RMS_{RA/DEC}$ $[arcsec]$ | $\chi^2_\nu$ |
|---|---|---|---|---|---|
| *Standard Model* (RT) $g(r) \propto r^{-k}$ | 2 | $5.5 \cdot 10^4$ | 107.0 | 0.75 / 0.88 | $2.5 \cdot 10^4$ |
| | 3 | $3.9 \cdot 10^4$ | 92.6 | 0.62 / 0.85 | $1.2 \cdot 10^4$ |
| | 4 | $3.1 \cdot 10^4$ | 81.8 | 0.55 / 0.81 | $6.5 \cdot 10^3$ |
| | 5 | $2.6 \cdot 10^4$ | 73.6 | 0.52 / 0.77 | $4.3 \cdot 10^3$ |
| | 6 | $2.3 \cdot 10^4$ | 67.2 | 0.51 / 0.74 | $3.4 \cdot 10^3$ |
| *Extended Standard Model* (RT) $g(r(t)) \propto r(t-DT)^{-k}$ | 2 | $5.2 \cdot 10^4$ | 87.9 | 0.71 / 0.72 | $2.4 \cdot 10^4$ |
| | 3 | $3.5 \cdot 10^4$ | 66.6 | 0.58 / 0.66 | $1.0 \cdot 10^4$ |
| | 4 | $2.4 \cdot 10^4$ | 50.5 | 0.53 / 0.60 | $4.8 \cdot 10^3$ |
| | 5 | $1.7 \cdot 10^4$ | 38.4 | 0.50 / 0.55 | $2.2 \cdot 10^3$ |
| | 6 | $1.1 \cdot 10^4$ | 29.4 | 0.49 / 0.52 | 914 |
| *Standard Model* (RTN) $g(r) \propto r^{-k}$ | 5 | $2.6 \cdot 10^4$ | 69.1 | 0.52 / 0.76 | $4.3 \cdot 10^3$ |
| | 6 | $2.4 \cdot 10^4$ | 53.3 | 0.49 / 0.69 | $3.3 \cdot 10^3$ |
| *Extended Standard Model* (RTN) $g(r) \propto r^{-k}$ | 5 | $1.3 \cdot 10^4$ | 88.9 | 0.59 / 0.87 | $1.7 \cdot 10^4$ |
| | 6 | $8.5 \cdot 10^3$ | 60.3 | 0.53 / 0.70 | 611 |
| *Rotating Jet Model* $g(r) \propto r^{-k}$ | 5 | $1.2 \cdot 10^3$ | 7.3 | 0.49 / 0.45 | 6.02 |
| *Linear stochastic model* $\Delta T_0 = 6\ days, C = 25$ | - | 105.8 | 0.5 | 0.48 / 0.43 | 0.39 |

The last explored model is the linear stochastic model. The *a priori* uncertainty for the degree-zero parameters $\vec{\alpha}_k$, normalized to 1 AU, were considered equal to the scale factor $A_i$ described above. Uncertainties at different heliocentric distances were then scaled according to the $r'^{-5}$ power law used to model the ESM and the RJM sublimation rate. Having no information on the derivative of the acceleration components, the *a priori* uncertainty for the degree-one coefficients $\vec{\beta}_k$ were obtained by dividing the acceleration uncertainty $\vec{\alpha}_k$ by the length of the stochastic interval $T_k$.

A series of filter runs were then performed by varying the minimum stochastic interval $T_0$, ranging between 1 and 15 days, and the constraint factor $C$, which was varied between 1 (i.e., no constraint) and 500. Low values of $T_0$ are required to capture short-scale outgassing variability around perihelion. However, as the number of stochastic intervals and estimated parameters increase, so does the risk of over-fitting and generating non-physical solutions. This effect is mainly caused by the range biases induced by the relative orbit discontinuities, which the filter tries to compensate for with artificially fast variations of the estimated NGAs. The constraint factor $C$ plays a crucial role in reducing the variability of estimated acceleration components between successive stochastic intervals, mitigating the effect of range biases. At the same time, high values of $C$ could suppress actual variations of the NGAs, reducing the scientific return of the analysis. A careful tradeoff of these two parameters was thus required for the ephemeris reconstruction. Figure 3 shows the $\chi_\nu^2$ values and the normalized residuals as a function of these two parameters.

A first selection of the suitable orbits was performed by keeping the solutions for which $\chi_\nu^2 < 0.9$, ensuring an overall goodness of fit. We can see from Figure 3 that several combinations of the free parameters satisfy this condition, except for solutions having $C > 300$. It should be noted that most of the solutions show a $\chi_\nu^2$ that is significantly smaller than one. This fact is mainly attributed to the uncertainty of the MPC astrometric measurement, whose sigma is conservatively assumed.

A further selection was performed by keeping the solutions for which the root square sum of the normalized residuals (with respect to the measurement uncertainty) is in the interval $\sqrt{\sum(r_i/\sigma_i)^2} \in [0.6, 0.9]$ for all the individual measurement types, namely 2-way range, ΔDOR, right ascension, and declination, ensuring that no measurement type dominates over the others. This condition is particularly relevant for the ΔDOR observables, which are highly sensible to the constraint factor, as shown in Figure 3, but whose residuals have a limited impact on the value of the $\chi_\nu^2$ statistics due to the limited number of measurements.

We can observe from Figure 3 that 27 solutions satisfy these conditions, with values of $T_0$ ranging between 1 and 15 days and constraint factors $C$ between 25 and 50. Out of these alternative orbits, the reference ephemeris solution, whose statistics are summarized in Table 3, was selected as the one with a minimum stochastic interval $T_0 = 6\ days$ and a constrain factor $C = 25$, representing the geometric center of the bundle of candidate orbits. The uncertainty for this reconstruction was obtained by multiplying its formal state covariance by a safety factor between 3 and 5, which ensured that all candidate solutions from Figure 3 fell within a 1σ distance from the reference solution and were thus statistically compatible.

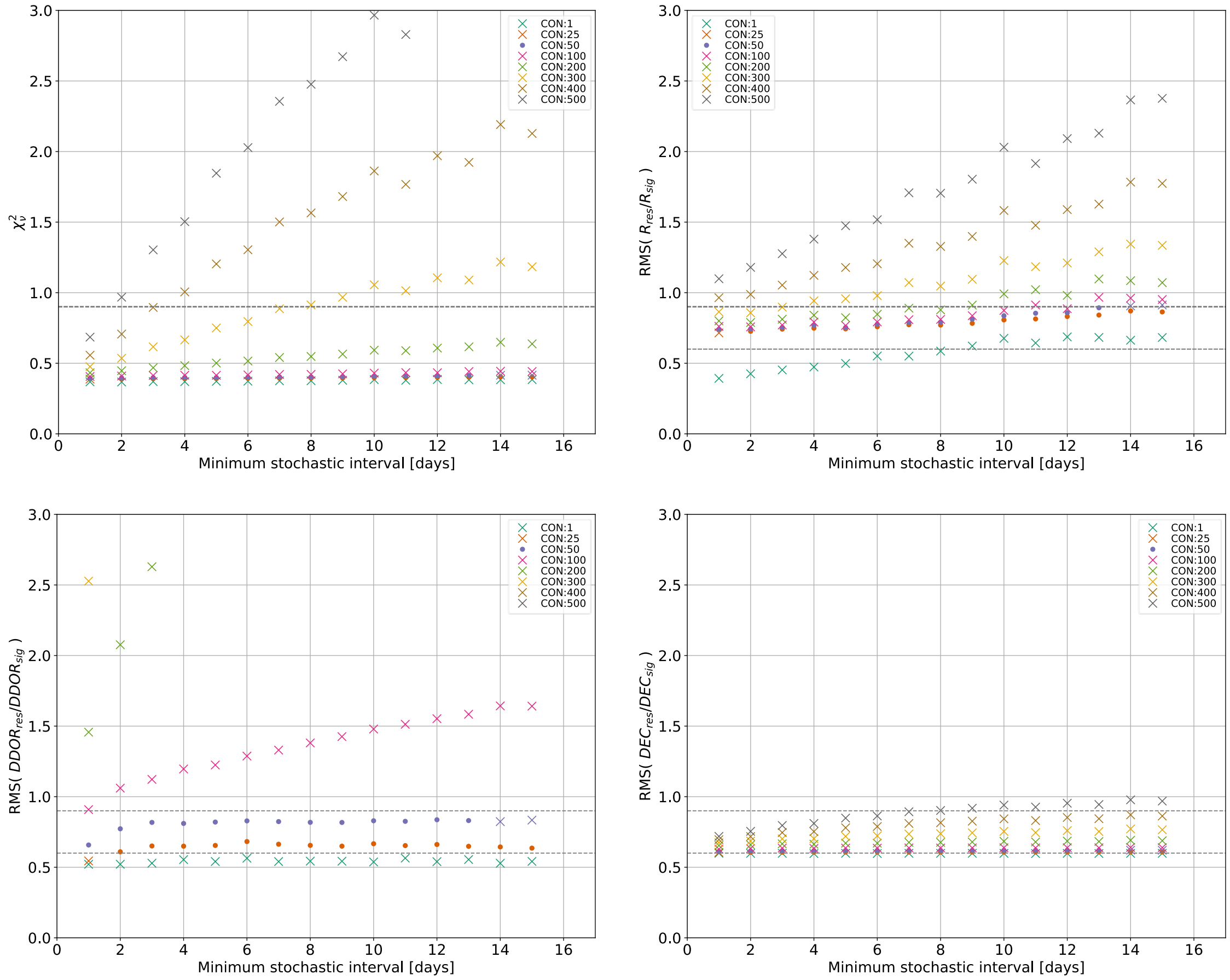

**Figure 3 Residuals' statistics for the LSM as a function of the minimum stochastic interval $T_0$ and of the constraint factor $C$. Dots: candidate ephemeris solutions; crosses: discarded solutions. Top left: reduced chi-square; top right: normalized range residuals; bottom left: normalized ΔDOR residuals; bottom right: normalized declination residuals (right ascension is not reported since all values satisfy the imposed condition).**

## Reference solution

Figure 4 shows the residuals for the reference ephemeris solution. During periods of proximity navigation, when the spacecraft's relative trajectory is accurately known, range residuals have zero mean and standard deviation in the order of a few meters. Conversely, range residuals degrade visibly when the spacecraft is far from the comet (e.g., around the perihelion or during the Far Approach Trajectory phase in July 2014), reaching values of hundreds of meters. This is particularly evident around perihelion, where we observe noise values consistent with the discontinuities of Rosetta's relative trajectory. Right ascension and declination residuals show root mean square values in the order of 0.15 and 0.5 arcseconds, respectively, for the JPL- and MPC-derived measurements, which are consistent with the values found in previous studies[8].

Although the observed residuals are far from showing a white noise dispersion, they are consistent with the proposed weighting scheme, as indicated by the weighted residuals in Figure 5, which always remain confined within a 3σ confidence interval, except for a small number of astrometric and range data points.

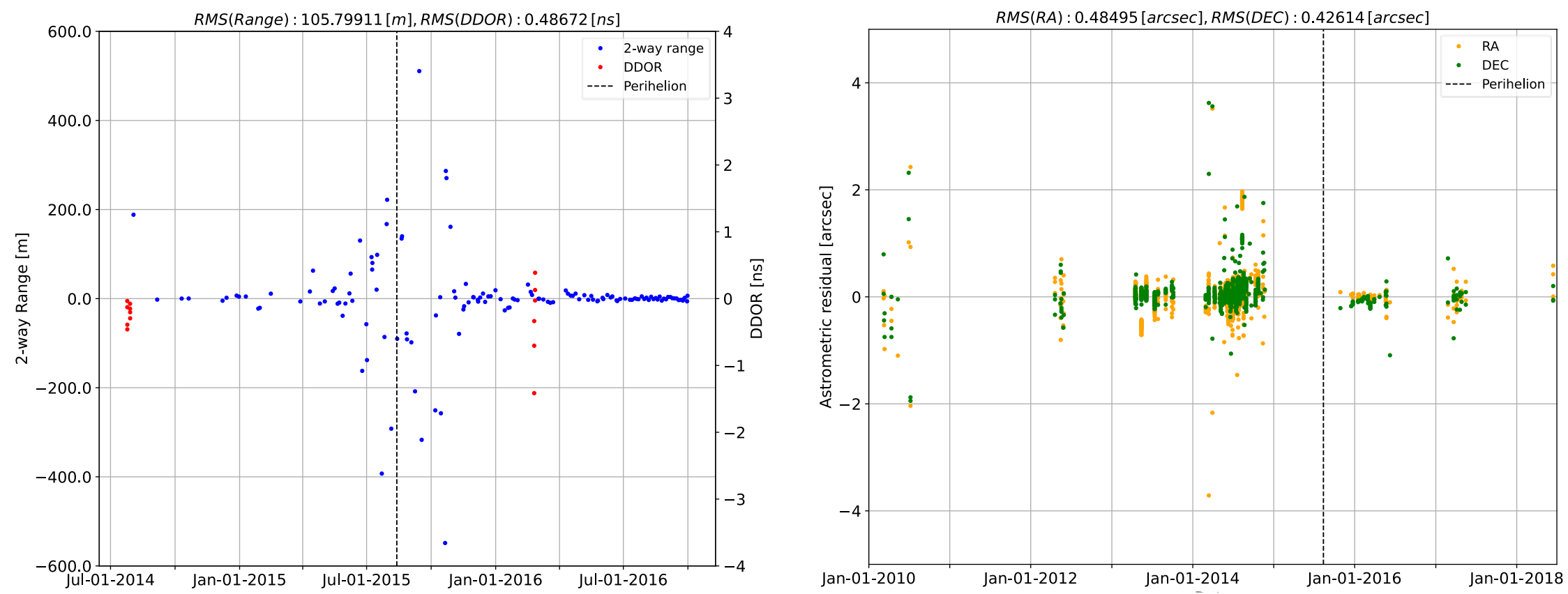

**Figure 4 Left: 2-way range and ΔDOR residuals for the reference ephemeris solution; right: right ascension and declination residuals for the reference ephemeris solution. Vertical dashed lines indicate the time perihelion passage.**

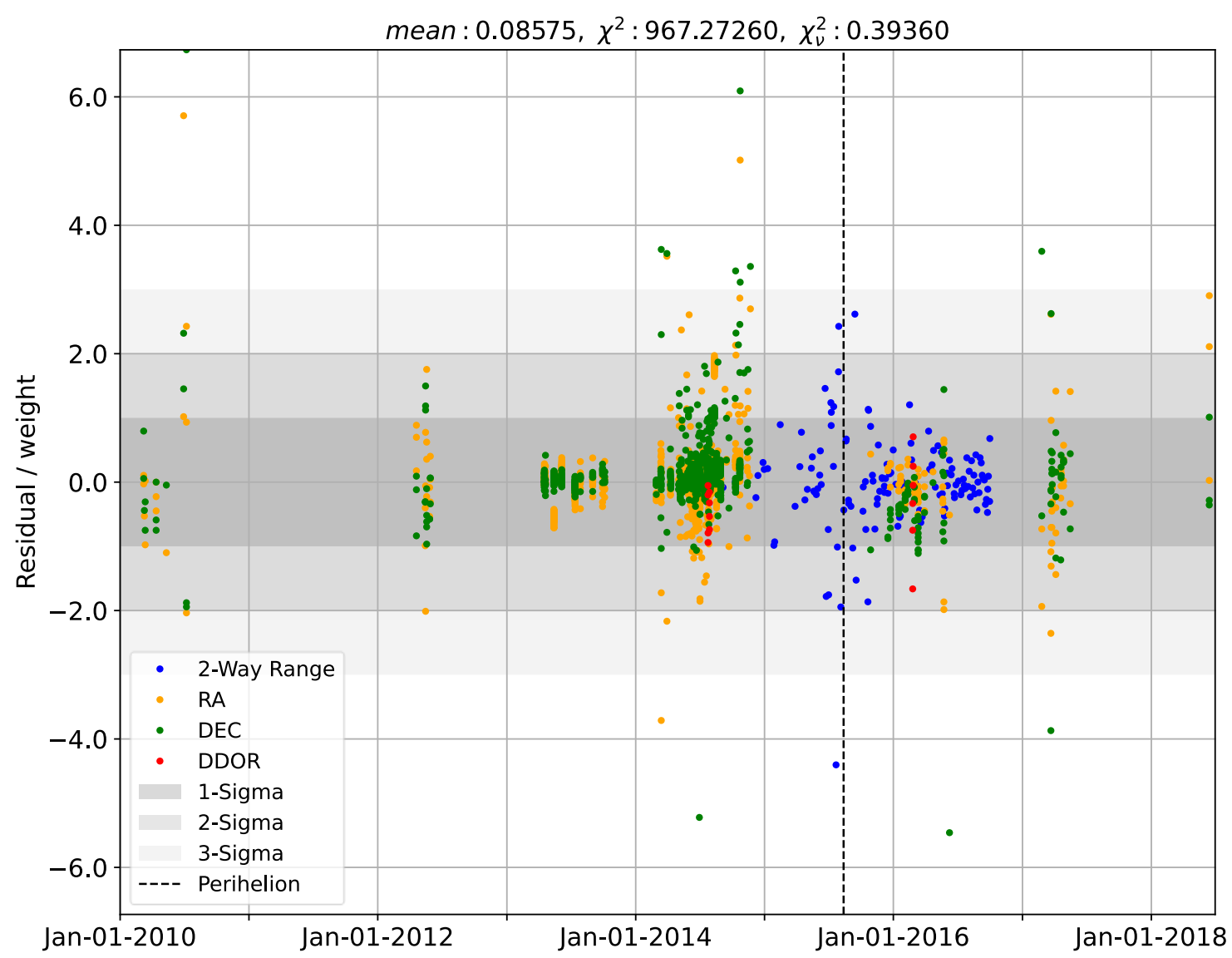

**Figure 5 Weighted measurement residuals for the reference ephemeris solution. The vertical dashed line indicates the perihelion. Horizontal grey stripes indicate the 1σ, 2σ, and 3σ confidence intervals.**

Figure 6 shows the estimated values and uncertainties for the stochastic NGA components in the orbital RTN frame. It can be seen that the acceleration is dominated by the radial component, which has a peak value of $a_R = (1.28 \pm 0.17) \cdot 10^{-8}\ m/s^2$ occurring roughly 15 days after perihelion. This value is nearly three times that of the normal component, whose peak value of $a_N = (0.52 \pm 0.20) \cdot 10^{-8}\ m/s^2$ occurs roughly 24 days after perihelion. These observed time offsets are compatible with the $\Delta T$ values estimated for the extended standard and rotating jet models. The tangential component is smaller in magnitude with respect to the other two, reaching peak absolute values of $a_T = (0.33 \pm 0.11) \cdot 10^{-8}\ m/s^2$ just before perihelion, and changes sign during the comet's active period.

By fitting a power law to the estimated acceleration magnitudes, we can see that the bundle of candidate solutions is best approximated by a function proportional to $r^{-6.4}$, steeper than the assumed water ice sublimation rate for the global models. However, the observed higher accelerations with respect to the best fit and the slight asymmetries around the perihelion suggest that this simple function of the heliocentric distance does not perfectly describe the overall outgassing rate. Seasonal variations of the local insolation and interactions between the dusty surface and the ice mantle have already been suggested as possible causes for time variations of the surface's active fraction (particularly in the South hemisphere) and might explain this sudden increase of the NGA magnitude around perihelion [2,35].

It should be noted that the uncertainties shown in Figure 6 already include a safety margin between 2 and 4 with respect to the estimated formal uncertainties of the respective accelerations, following the same approach applied to the comet's state. This way, all the candidate solutions for the linear stochastic model, displayed as dashed lines in Figure 6, fall within a 1σ uncertainty from the reference one, making them statistically compatible.

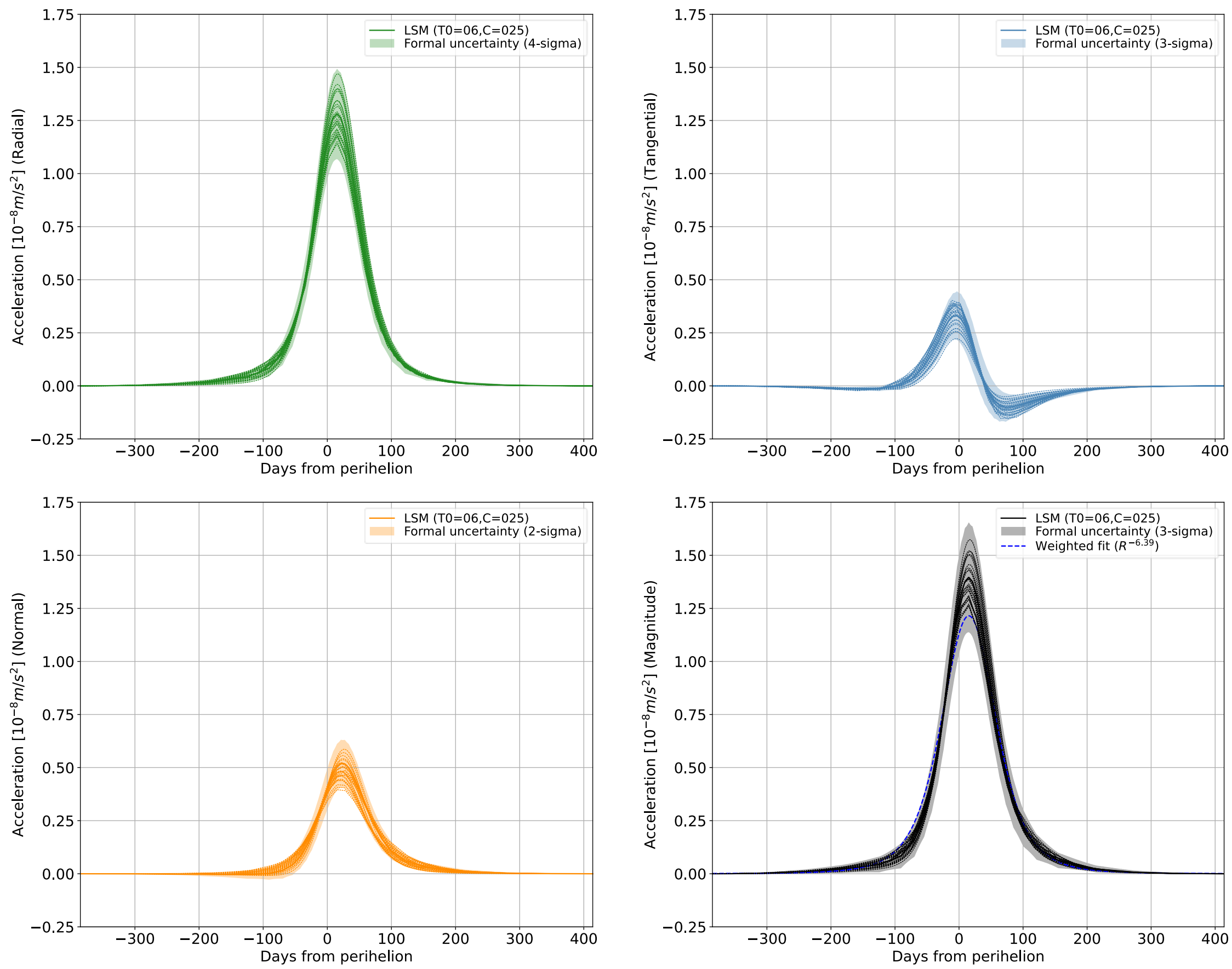

**Figure 6 Estimated non-gravitational acceleration components for the reference ephemeris solution (continuous line) and for alternative candidate solutions from Figure 3 (dashed lines). Top left: radial component; top right: tangential component; bottom left: normal component; bottom right: acceleration magnitude. Shaded areas indicate the acceleration uncertainties for the reference solution, which correspond to the estimated formal uncertainties multiplied by a safety factor between 2 and 4.**

For a complete characterization of the reference ephemeris reconstruction, it is interesting to compare this solution with the operational trajectory of Rosetta's navigation team. Figure 7 shows the position differences between the alternative ephemeris reconstructions in a reference frame, which is defined as follows: the first axis is the Earth-comet line of sight, the third axis is the comet orbit-normal, and the second axis completes the right-handed orthogonal frame.

Most of the information content for the 2-way range is provided along the Earth-comet direction, as shown by the corresponding position uncertainties in Figure 7, ranging between roughly 10-20 m during proximity navigation and 200-300 m after perihelion. These uncertainties have an order magnitude similar to the ones of the discontinuities in the operational trajectory, which appear as zero-mean oscillations around the estimated solution. An exception to the zero-mean appearance of these discontinuities is represented by the February 2015 solar conjunction, where we observe an offset that is compatible with the value for the range bias induced by solar plasma of $\Delta\rho_{SP} = (117 \pm 12)\ m$. This was expected since the operational ephemeris solution used for navigation did not include the effect of solar plasma within the observational model.

The position uncertainty along the comet orbit-normal direction is roughly three orders of magnitude higher than the one along the Earth-comet direction due to the limited information content provided by the 2-way range for the low declination values observed during Rosetta's proximity phase. Two minima are observed in the orbit-normal uncertainty in July 2014 and February 2016 in correspondence with the ΔDOR measurements, which provide a tight constraint to the inclination of the comet's orbital plane. Position uncertainty along the cross-axis direction represents an intermediate case with respect to the previous two, with values ranging from roughly 10 km to a maximum of 25 km around perihelion.

Earth-based astrometry provides limited information on the comet's plane of the sky position around perihelion. This is primarily due to their lower accuracy with respect to spacecraft-based radiometric observables. Furthermore, observational biases induced by the comet's tail limited the number of measurements that could be included within the filter. However, it should be noted that the MPC astrometry used in the OD process extends well beyond Rosetta's proximity operations. Specifically, the data covers an interval between March 2010 and July 2018, representing a complete heliocentric orbit of 67P (2012-2018) and the outbound portion (i.e., post-perihelion) of the previous orbit, which is expected to have minimal outgassing activity ($r_{\odot} > 3.5$). By constraining the comet's plane of the sky position at the boundaries of the selected interval, these measurements increase the robustness of the solution and reduce the uncertainty in the normal and cross components during Rosetta's proximity phase.

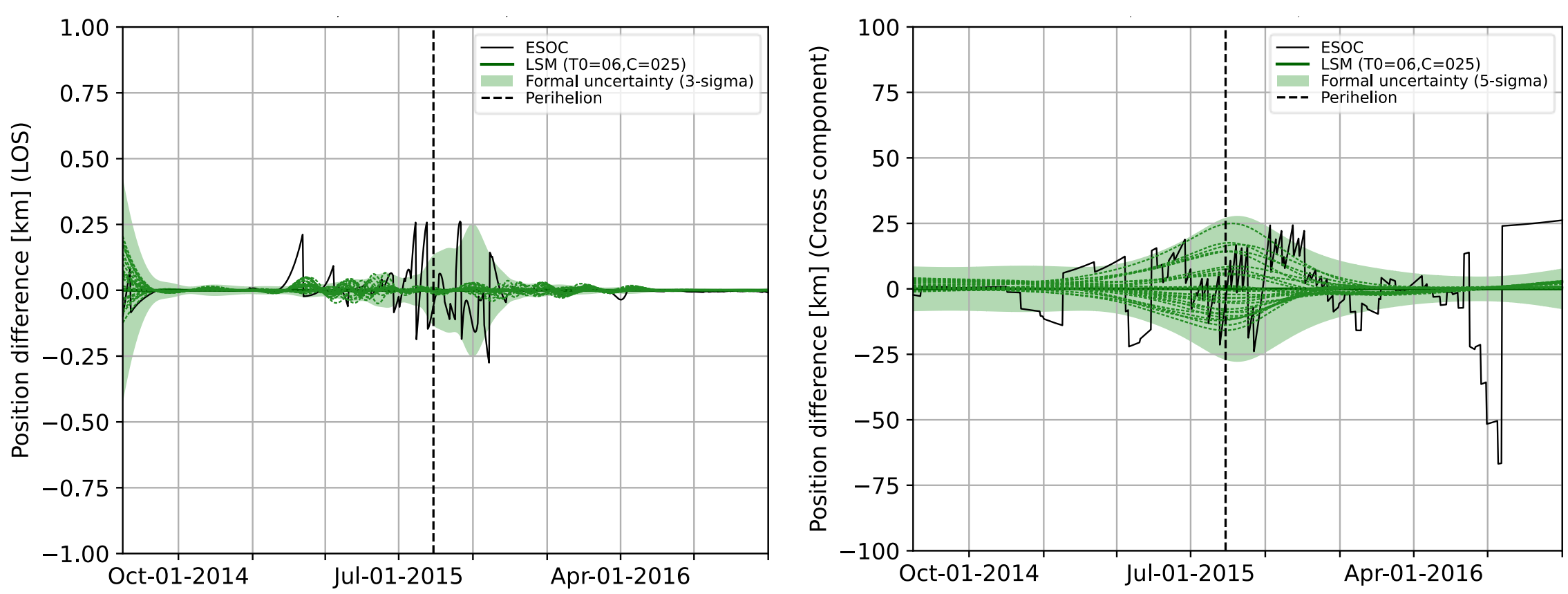

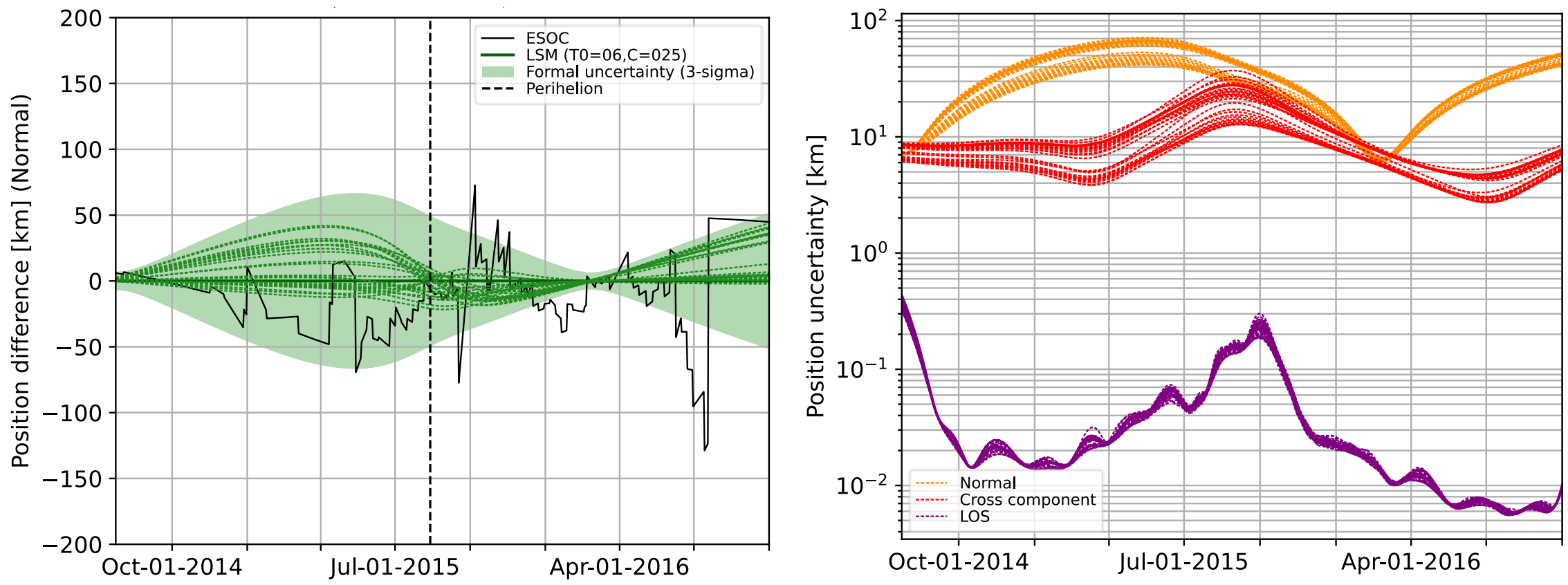

**Figure 7 Comparison between the reconstructed ephemeris solution (thick continuous line), the alternative candidate orbits (dashed lines), and the operational trajectory from the navigation team (solid black line). Top left: position differences in the Earth-comet direction; top right: position differences in the cross-component direction; bottom left: position differences in the comet orbit-normal direction; bottom right: position uncertainties as a function of time using a logarithmic scale.**

## CONCLUSION

This work provides a continuous and accurate ephemeris reconstruction for comet 67P/Churyumov-Gerasimenko during Rosetta's proximity phase between July 2014 and September 2016.

Using the spacecraft to comet relative trajectory from the navigation team, Earth-based radiometric and astrometric measurements were used to estimate the comet state, along with key physical and observational parameters, most notably the non-gravitational accelerations induced by outgassing of sub-surface volatiles. Several estimations were performed by varying the dynamical model used to represent the non-gravitational motion of the comet, leading to the following results:

a) Simple global models such as the extended standard model and the rotating jet model are inadequate for fitting the radiometric measurements around the perihelion, where short-term variations of the outgassing activity are observed.

b) Empirical stochastic models using polynomial functions to represent the non-gravitational accelerations can produce acceptable fits. Specifically, the reconstructed ephemeris solution, which uses piecewise-continuous linear accelerations, provides maximum position uncertainties around perihelion near 10 km, 30 km, and 80 km in the orbital radial, tangential, and normal directions.

c) The selected model allowed the production of an agnostic estimation of the non-gravitational acceleration, whose module is best fit by a $r^{-6}$ power law of the heliocentric distance and reaches its peak value roughly 15 days after perihelion. Relying on limited assumptions on the involved physical processes, the estimated acceleration components may represent valuable input for further investigations involving high-fidelity thermo-physical models of the surface's activity.

It should also be noted that the estimation of an accurate heliocentric trajectory for 67P/CG is strictly coupled with the quality of the relative orbit reconstruction for Rosetta. Specifically, the lack of knowledge of the relative orbit uncertainty has proven to be a limiting factor for the definition of a suitable weighting scheme of the radiometric measurements, for which an empirical function of the cometocentric distance was used. Moreover, the non-gravitational accelerations estimated with the linear stochastic model were shown to be highly sensitive to the observational biases induced by the discontinuities in Rosetta's relative orbit. A careful selection of the stochastic interval length and the constraints between successive acceleration values was needed to mitigate the effects of these observational biases. Future work may thus include re-analyzing Doppler and optical data to reconstruct Rosetta's relative trajectory, which might significantly improve the accuracy of the comet's ephemeris.

## ACKNOWLEDGMENTS

This research was supported by the International Space Science Institute (ISSI) in Bern through ISSI International Team project #547 (Understanding the Activity of Comets Through 67P's Dynamics)."

RLM, MZ, and PT wish to acknowledge Caltech and the Jet Propulsion Laboratory for granting the University of Bologna a license to an executable version of MONTE Project Edition S/W.

## APPENDIX A: WEIGHTS VALIDATION

After validation of the orbit determination setup, a pass-through was performed, which consists in computing the expected observables using as input the *a priori* trajectories from ESOC FD and comparing them with the measured observables. This process was repeated for monthly arcs covering the whole mission duration to highlight inconsistencies in the processed datasets and perform manual removal of the outliers.

Figure 8 shows the pre-fit range residuals for July 2015, corresponding to a period of intense surface activity. If we look at the geocentric range discontinuities for comet 67P/CG (see Figure 1), we would expect to find at least one major discontinuity in the Rosetta pre-fit residuals, with an order of magnitude of roughly 800 m for the two-way range. The expected discontinuity is not present in the pre-fit residuals of Figure 8, where only a small 6 m gap is observed at the corresponding time tag. This supports the assumption that, during the operational OD reconstruction, discontinuities in the Earth-comet LOS direction have been absorbed by equal and opposite discontinuities in the relative orbit of Rosetta around the comet. As pointed out above, this also means that the uncertainty in the relative orbit reconstruction for the corresponding interval should have at least the same order of magnitude of the one-way range discontinuities from Figure 1 (if not higher). This observation was critical during the definition of the weighting scheme since it provided a lower limit for the relative orbit uncertainty during the period of maximum comet activity, where no other indication is available from the literature.

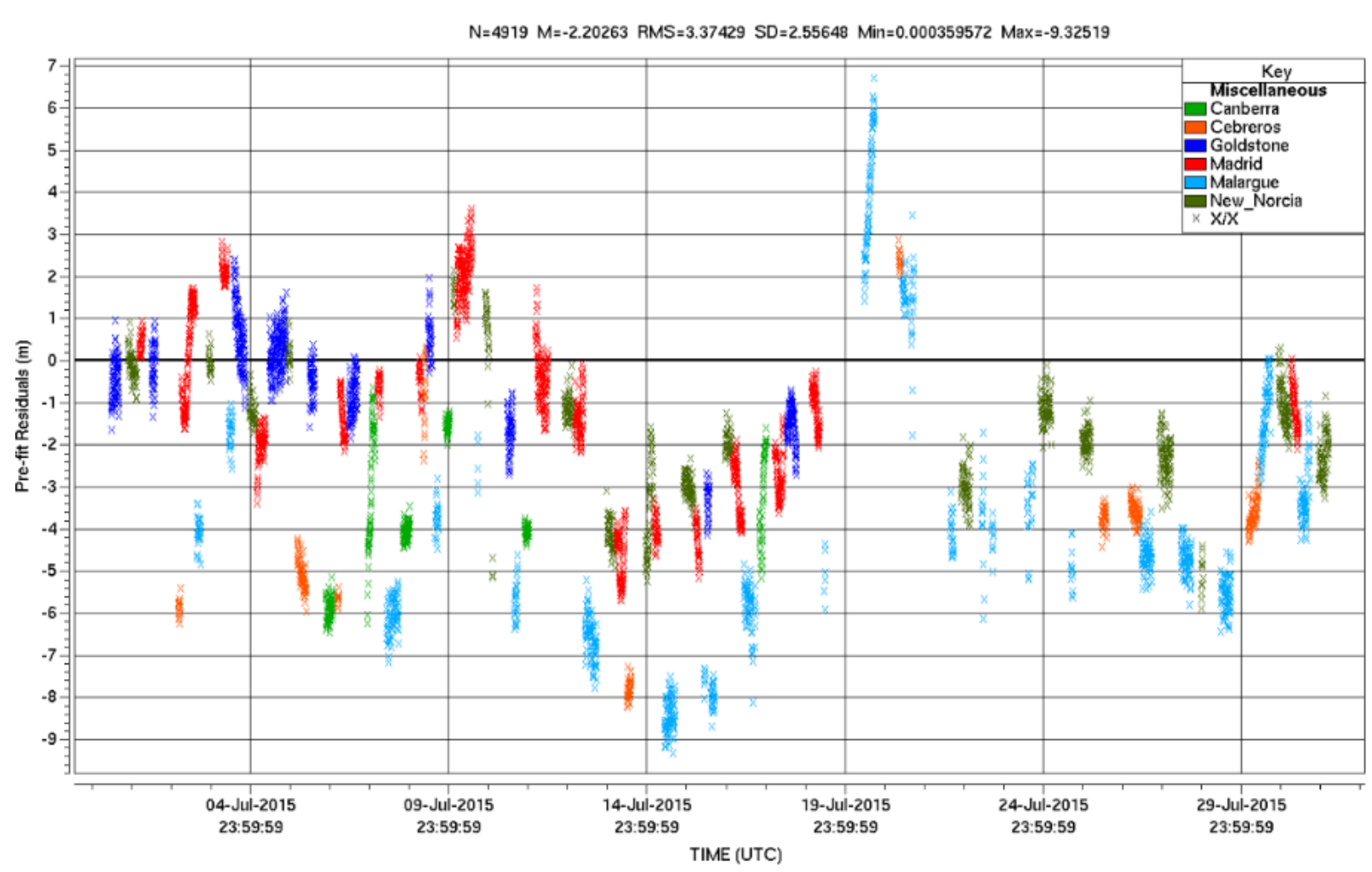

**Figure 8 Rosetta pre-fit range residuals for July 2015 using the ESOC FD kernels**

## APPENDIX B: SUPPLEMENTARY MATERIAL

Supplementary material containing the SPICE kernel for the reconstructed ephemeris of 67P and the estimated non-gravitational accelerations is available on Zenodo[36]